\input harvmac
\overfullrule=0pt
\parindent 25pt
\tolerance=10000
\input epsf

\newcount\figno
\figno=0
\def\fig#1#2#3{
\par\begingroup\parindent=0pt\leftskip=1cm\rightskip=1cm\parindent=0pt
\baselineskip=11pt
\global\advance\figno by 1
\midinsert
\epsfxsize=#3
\centerline{\epsfbox{#2}}
\vskip 12pt
{\bf Fig.\ \the\figno: } #1\par
\endinsert\endgroup\par
}
\def\figlabel#1{\xdef#1{\the\figno}}
\def\encadremath#1{\vbox{\hrule\hbox{\vrule\kern8pt\vbox{\kern8pt
\hbox{$\displaystyle #1$}\kern8pt}
\kern8pt\vrule}\hrule}}

 \def\ep{{\epsilon}}

 \def\frac#1#2{{#1\over #2}}

 \def\s{\sqrt}

 \def\al{\alpha'}
 \def\de{\partial}

 \def\lr{\leftrightarrow}
 \def\f {\frac}
 \def\ti{\tilde}
 \def\ap{\alpha}

 \def\la{\langle}
 \def\lb{\rangle}
 \def\ep{\epsilon}

\lref\Nstring{ M.~Ademollo {\it et al.}, ``Dual String With U(1)
Color Symmetry,'' Nucl.\ Phys.\ B {\bf 111}, 77 (1976).
%%CITATION = NUPHA,B111,77;%%
}

\lref\OVH{H.~Ooguri and C.~Vafa, ``N=2 heterotic strings,'' Nucl.\
Phys.\ B {\bf 367}, 83 (1991).
%%CITATION = NUPHA,B367,83;%%
}

\lref\OV{H.~Ooguri and C.~Vafa, ``Two-Dimensional Black Hole and
Singularities of CY Manifolds,'' Nucl.\ Phys.\ B {\bf 463}, 55
(1996) [arXiv:hep-th/9511164].
%%CITATION = HEP-TH 9511164;%%
}

\lref\KuMa{D.~Kutasov and E.~J.~Martinec, ``M-branes and N = 2
strings,'' Class.\ Quant.\ Grav.\  {\bf 14}, 2483 (1997)
[arXiv:hep-th/9612102].
%%CITATION = HEP-TH 9612102;%%
}

\lref\OL{ M.~J.~O'Loughlin and S.~Randjbar-Daemi, ``AdS(3) x R as
a target space for the (2,1) string theory,'' Nucl.\ Phys.\ B {\bf
543}, 170 (1999) [arXiv:hep-th/9807208].
%%CITATION = HEP-TH 9807208;%%
}

\lref\Nopen{N.~Marcus, ``The N=2 open string,'' Nucl.\ Phys.\ B
{\bf 387}, 263 (1992) [arXiv:hep-th/9207024].
%%CITATION = HEP-TH 9207024;%%
}

\lref\Nreview{N.~Marcus, ``A Tour through N=2 strings,''
arXiv:hep-th/9211059.
%%CITATION = HEP-TH 9211059;%%
}

\lref\GiSe{A.~Giveon and A.~Sever, ``Strings in a 2-d extremal
black hole,'' JHEP {\bf 0502}, 065 (2005) [arXiv:hep-th/0412294].
%%CITATION = HEP-TH 0412294;%%
}

\lref\COY{Y.~K.~Cheung, Y.~Oz and Z.~Yin, ``Families of N = 2
strings,'' JHEP {\bf 0311}, 026 (2003) [arXiv:hep-th/0211147].
%%CITATION = HEP-TH 0211147;%%
}

\lref\GOS{D.~Gluck, Y.~Oz and T.~Sakai, ``The effective action and
geometry of closed N = 2 strings,'' JHEP {\bf 0307}, 007 (2003)
[arXiv:hep-th/0304103].
%%CITATION = HEP-TH 0304103;%%
}

\lref\GOSD{D.~Gluck, Y.~Oz and T.~Sakai, ``D-branes in N = 2
strings,'' JHEP {\bf 0308}, 055 (2003) [arXiv:hep-th/0306112].
%%CITATION = HEP-TH 0306112;%%
}

\lref\Hull{C.~M.~Hull, ``The geometry of N = 2 strings with
torsion,'' Phys.\ Lett.\ B {\bf 387}, 497 (1996)
[arXiv:hep-th/9606190].
%%CITATION = HEP-TH 9606190;%%
}

\lref\Hos{K.~Hosomichi, ``N = 2 Liouville theory with boundary,''
arXiv:hep-th/0408172.
%%CITATION = HEP-TH 0408172;%%
}

\lref\OVN{H.~Ooguri and C.~Vafa, ``Selfduality And N=2 String
Magic,'' Mod.\ Phys.\ Lett.\ A {\bf 5}, 1389 (1990);
%%CITATION = MPLAE,A5,1389;%%
``Geometry of N=2 strings,'' Nucl.\ Phys.\ B {\bf 361}, 469
(1991).
%%CITATION = NUPHA,B361,469;%%
}

\lref\BV{N.~Berkovits and C.~Vafa, ``N=4 topological strings,''
Nucl.\ Phys.\ B {\bf 433}, 123 (1995) [arXiv:hep-th/9407190].
%%CITATION = HEP-TH 9407190;%%
}

\lref\TLD{ T.~Takayanagi, ``Matrix model and time-like linear
dilaton matter,'' JHEP {\bf 0412}, 071 (2004)
[arXiv:hep-th/0411019].
%%CITATION = HEP-TH 0411019;%%
}

\lref\DHK{
J.~Distler, Z.~Hlousek and H.~Kawai,
``Superliouville Theory As A Two-Dimensional, Superconformal Supergravity
Theory,''
Int.\ J.\ Mod.\ Phys.\ A {\bf 5}, 391 (1990).
%%CITATION = IMPAE,A5,391;%%
}

\lref\ABK{
I.~Antoniadis, C.~Bachas and C.~Kounnas,
``N=2 Superliouville And Noncritical Strings,''
Phys.\ Lett.\ B {\bf 242}, 185 (1990).
%%CITATION = PHLTA,B242,185;%%
}

\lref\TMV{T.~Takayanagi, ``$c < 1$ string from two dimensional
black holes,'' arXiv:hep-th/0503237.
%%CITATION = HEP-TH 0503237;%%
}

\lref\AAD{E.~Abdalla, M.~C.~B.~Abdalla and D.~Dalmazi, ``On the
amplitudes for noncritical N=2 superstrings,'' Phys.\ Lett.\ B
{\bf 291}, 32 (1992) [arXiv:hep-th/9206054];
%%CITATION = HEP-TH 9206054;%%
D.~Dalmazi, ``Tree amplitudes in noncritical N=2 strings,''
arXiv:hep-th/9209056.
%%CITATION = HEP-TH 9209056;%%
}

\lref\KKL{E.~Kiritsis, C.~Kounnas and D.~Lust, ``A Large class of
new gravitational and axionic backgrounds for four-dimensional
superstrings,'' Int.\ J.\ Mod.\ Phys.\ A {\bf 9}, 1361 (1994)
[arXiv:hep-th/9308124].
%%CITATION = HEP-TH 9308124;%%
}

\lref\tdsa{ D.~J.~Gross and N.~Miljkovic, ``A Nonperturbative
Solution Of D = 1 String Theory,'' Phys.\ Lett.\ B {\bf 238}, 217
(1990);
%%CITATION = PHLTA,B238,217;%%
E.~Brezin, V.~A.~Kazakov and A.~B.~Zamolodchikov, ``Scaling
Violation In A Field Theory Of Closed Strings In One Physical
Dimension,'' Nucl.\ Phys.\ B {\bf 338}, 673 (1990);
%%CITATION = NUPHA,B338,673;%%
P.~Ginsparg and J.~Zinn-Justin, ``2-D Gravity + 1-D Matter,''
Phys.\ Lett.\ B {\bf 240}, 333 (1990).
%%CITATION = PHLTA,B240,333;%%
}

\lref\DK{P.~Di Francesco and D.~Kutasov, ``Correlation functions
in 2-D string theory,'' Phys.\ Lett.\ B {\bf 261}, 385 (1991);
%%CITATION = PHLTA,B261,385;%%
``World sheet and space-time physics in two-dimensional
(Super)string theory,'' Nucl.\ Phys.\ B {\bf 375}, 119 (1992)
[arXiv:hep-th/9109005].
%%CITATION = HEP-TH 9109005;%%
}

\lref\MV{ J.~McGreevy and H.~Verlinde, ``Strings from tachyons:
The c = 1 matrix reloaded,'' JHEP {\bf 0312}, 054 (2003)
[arXiv:hep-th/0304224];
%%CITATION = HEP-TH 0304224;%%
J.~McGreevy, J.~Teschner and H.~L.~Verlinde, ``Classical and
quantum D-branes in 2D string theory,'' JHEP {\bf 0401}, 039
(2004) [arXiv:hep-th/0305194].
%%CITATION = HEP-TH 0305194;%%
}

\lref\KMS{ I.~R.~Klebanov, J.~Maldacena and N.~Seiberg, ``D-brane
decay in two-dimensional string theory,'' JHEP {\bf 0307}, 045
(2003) [arXiv:hep-th/0305159].
%%CITATION = HEP-TH 0305159;%%
}

\lref\TT{T.~Takayanagi and N.~Toumbas, ``A matrix model dual of
type 0B string theory in two dimensions,'' JHEP {\bf 0307}, 064
(2003) [arXiv:hep-th/0307083];
%%CITATION = HEP-TH 0307083;%%

S.~Gukov, T.~Takayanagi and N.~Toumbas, ``Flux backgrounds in 2D
string theory,'' JHEP {\bf 0403}, 017 (2004)
[arXiv:hep-th/0312208].
%%CITATION = HEP-TH 0312208;%%
}

\lref\six{M.~R.~Douglas, I.~R.~Klebanov, D.~Kutasov, J.~Maldacena,
E.~Martinec and N.~Seiberg, ``A new hat for the c = 1 matrix
model,'' [arXiv:hep-th/0307195];
%%CITATION = HEP-TH 0307195;%%

J.~Maldacena and N.~Seiberg, ``Flux-vacua in two dimensional
string theory,''  arXiv:hep-th/0506141.
%%CITATION = HEP-TH 0506141;%%
}

\lref\COL{H.~Dorn and H.~J.~Otto, ``Two and three point functions
in Liouville theory,'' Nucl.\ Phys.\ B {\bf 429}, 375 (1994)
[arXiv:hep-th/9403141];
%%CITATION = HEP-TH 9403141;%%

A.~B.~Zamolodchikov and A.~B.~Zamolodchikov, ``Structure constants
and conformal bootstrap in Liouville field theory,'' Nucl.\ Phys.\
B {\bf 477}, 577 (1996) [arXiv:hep-th/9506136];
%%CITATION = HEP-TH 9506136;%%

Y.~Nakayama, ``Liouville field theory: A decade after the
revolution,'' arXiv:hep-th/0402009.
%%CITATION = HEP-TH 0402009;%%
}

\lref\ST{ A.~Strominger and T.~Takayanagi, ``Correlators in
timelike bulk Liouville theory,'' Adv.\ Theor.\ Math.\ Phys.\
{\bf 7}, 369 (2003) [arXiv:hep-th/0303221].
%%CITATION = HEP-TH 0303221;%%
}

\lref\SC{V.~Schomerus, ``Rolling tachyons from Liouville theory,''
JHEP {\bf 0311}, 043 (2003) [arXiv:hep-th/0306026].
%%CITATION = HEP-TH 0306026;%%
}

%%%

\lref\HT{ Y.~Hikida and T.~Takayanagi, ``On solvable
time-dependent model and rolling closed string tachyon,''
arXiv:hep-th/0408124.
%%CITATION = HEP-TH 0408124;%%
}

\lref\ES{T.~Eguchi and Y.~Sugawara, ``Modular bootstrap for
boundary N = 2 Liouville theory,'' JHEP {\bf 0401}, 025 (2004)
[arXiv:hep-th/0311141];
%%CITATION = HEP-TH 0311141;%%
T.~Eguchi, ``Modular bootstrap of boundary N = 2 Liouville
theory,'' Comptes Rendus Physique {\bf 6}, 209 (2005)
[arXiv:hep-th/0409266].
%%CITATION = HEP-TH 0409266;%%
}

\lref\AR{C.~Ahn, M.~Stanishkov and M.~Yamamoto, ``One-point
functions of N = 2 super-Liouville theory with boundary,'' Nucl.\
Phys.\ B {\bf 683}, 177 (2004) [arXiv:hep-th/0311169].
%%CITATION = HEP-TH 0311169;%%
}

\lref\IPT{D.~Israel, A.~Pakman and J.~Troost, ``D-branes in N = 2
Liouville theory and its mirror,'' Nucl.\ Phys.\ B {\bf 710}, 529
(2005) [arXiv:hep-th/0405259].
%%CITATION = HEP-TH 0405259;%%
}

\lref\GK{A.~Giveon and D.~Kutasov, ``Comments on double scaled
little string theory,'' JHEP {\bf 0001}, 023 (2000)
[arXiv:hep-th/9911039].
%%CITATION = HEP-TH 9911039;%%
}

\lref\Sen{A.~Sen, ``Open-closed duality: Lessons from matrix
model,'' Mod.\ Phys.\ Lett.\ A {\bf 19}, 841 (2004)
[arXiv:hep-th/0308068].
%%CITATION = HEP-TH 0308068;%%
}

\lref\TaTe{T.~Takayanagi and S.~Terashima, ``c = 1 matrix model
from string field theory,'' arXiv:hep-th/0503184.
%%CITATION = HEP-TH 0503184;%%
}

\lref\HK{K.~Hori and A.~Kapustin, ``Duality of the fermionic 2d
black hole and N = 2 Liouville theory as mirror symmetry,'' JHEP
{\bf 0108}, 045 (2001) [arXiv:hep-th/0104202].
%%CITATION = HEP-TH 0104202;%%
}

\lref\HoA{T.~Fukuda and K.~Hosomichi, ``Three-point functions in
sine-Liouville theory,'' JHEP {\bf 0109}, 003 (2001)
[arXiv:hep-th/0105217].
%%CITATION = HEP-TH 0105217;%%
}

%%%%

\lref\AH{O.~Aharony, B.~Fiol, D.~Kutasov and D.~A.~Sahakyan,
``Little string theory and heterotic/type II duality,'' Nucl.\
Phys.\ B {\bf 679}, 3 (2004) [arXiv:hep-th/0310197].
%%CITATION = HEP-TH 0310197;%%
}

\lref\EY{T.~Eguchi and S.~K.~Yang, ``N=2 Superconformal Models As
Topological Field Theories,'' Mod.\ Phys.\ Lett.\ A {\bf 5}, 1693
(1990).
%%CITATION = MPLAE,A5,1693;%%
}

\lref\KPS{ A.~Konechny, A.~Parnachev and D.~A.~Sahakyan, ``The
Ground Ring of N=2 Minimal String Theory,'' arXiv:hep-th/0507002.
%%CITATION = HEP-TH 0507002;%%
}

\lref\Gir{G.~E.~Giribet and D.~E.~Lopez-Fogliani, ``Remarks on
free field realization of SL(2,R)k/U(1) x U(1) WZNW model,'' JHEP
{\bf 0406}, 026 (2004) [arXiv:hep-th/0404231].
%%CITATION = HEP-TH 0404231;%%
}

\lref\Jev{A.~Jevicki, M.~Mihailescu and J.~P.~Nunes, ``Large N
field theory of N = 2 strings and selfdual gravity,'' Chaos
Solitons Fractals {\bf 10}, 385 (1999) [arXiv:hep-th/9804206];
%%CITATION = HEP-TH 9804206;%%
``Large N WZW field theory of N = 2 strings,'' Phys.\ Lett.\ B
{\bf 416}, 334 (1998) [arXiv:hep-th/9706223].
%%CITATION = HEP-TH 9706223;%%
}

\lref\ADE{I.~K.~Kostov, ``Gauge invariant matrix model for the
A-D-E closed strings,'' Phys.\ Lett.\ B {\bf 297}, 74 (1992)
[arXiv:hep-th/9208053].
%%CITATION = HEP-TH 9208053;%%
}

\baselineskip 18pt plus 2pt minus 2pt

\Title{\vbox{\baselineskip12pt
\hbox{hep-th/0507065}\hbox{HUTP-05/A0034}
  }}
{\vbox{\centerline{Notes on S-Matrix of Non-critical $N=2$
String}}} \centerline{Tadashi Takayanagi\foot{e-mail:
takayana@bose.harvard.edu}}

\medskip\centerline{ \it Jefferson Physical Laboratory}
\centerline{\it Harvard University} \centerline{\it Cambridge, MA
02138, USA}

\vskip .5in \centerline{\bf Abstract}

In this paper we discuss the scattering S-matrix of non-critical
$N=2$ string at tree level. First we consider the $\hat{c}<1$
string defined by combining the $N=2$ time-like linear dilaton
SCFT with the $N=2$ Liouville theory. We compute three particle
scattering amplitudes explicitly and find that they are actually
vanishing. We also find an evidence that this is true for higher
amplitudes. Next we analyze another $\hat{c}<1$ string obtained
from the $N=2$ time-like Liouville theory, which is closely
related to the $N=2$ minimal string. In this case, we find a
non-trivial expression for the three point functions. When we
consider only chiral primaries, the amplitudes are very similar to
those in the $(1,n)$ non-critical bosonic string.

\noblackbox

\Date{July, 2005}

%\listtoc
\writetoc

\newsec{Introduction and Summary}

The $N=2$ fermionic string \Nstring\ (for a review see \Nreview)
is described by the most supersymmetric world-sheet among
physically interesting string theories which have positive
critical dimensions. Interestingly, its critical dimension is four
because the ghost anomaly cancelation requires $\hat{c}=2$. When
we consider the ordinary flat spacetime, there is only one
massless physical state and its spacetime theory is described by
the self-dual gravity \OVN. In this case, however, the $N=2$
world-sheet supersymmetry requires two time directions.

In spite of this signature problem, we can still believe that
$N=2$ string theory provides the most beautiful and simple model
at least from theoretical viewpoints. Indeed, the spectrum of
$N=2$ string in the four dimensional flat spacetime $R^{2,2}$ is
very similar to those in two dimensional string theories in
bosonic or type 0 string. These $N=0,1$ non-critical string
theories can be non-perturbatively solvable by the matrix model
descriptions \tdsa \TT \six\ via the open-closed duality \MV \KMS
\Sen \TaTe. Therefore, it will be very intriguing if we can
understand better a non-perturbative formulation of $N=2$ string
via an analogous open-closed duality.

Motivated by this, we would like to consider non-critical $N=2$
string models. They are defined by the $N=2$ Liouville theory plus
a matter $N=2$ SCFT with a central charge $\hat{c}<1$, such that
the total central charge is $\hat{c}=2$ \DHK \ABK \KKL \AAD \OV.
Indeed, our knowledge of such models is very little compared with
that of $N=0,1$ string. There is no known matrix model like
description though they are expected to be integrable. Recent
discussions can be found in \AH \KPS.

In this paper we first consider the $N=2$ $\hat{c}<1$ string whose
matter part is the time-like linear dilaton theory\foot{Linear
dilaton backgrounds in Heterotic $N=2$ string \OVH \ may also have
interesting properties as pointed out in \GiSe (see also \KuMa
\OL).}. Physically, this model describes a three dimensional
spacetime (one time, one space and one null direction) with a
dynamical massless scalar field. Also it can be regarded as a
$N=2$ string analogue of the $c<1$ deformation of the two
dimensional string whose matrix model dual was constructed
recently in \TLD. Notice that we can regard the ordinary $N=2$
string on $R^{2,2}$ as a particular $\hat{c}=1$ model in this
sense.

A basic dynamical property in these $N=2$ string models will be
the scattering amplitudes. We explicitly compute three particle
scattering S-matrix at tree level in the equivalent $N=4$
topological string formulation \BV , employing the recent
progresses on three point functions in the $N=2$ Liouville theory
\Hos\ or equivalent $N=2$ $SL(2,R)/U(1)$ coset \GK. In the end, we
find a very simple result that the three particle scattering
amplitudes are all zero in this $N=2$ $\hat{c}<1$ string theories.
There is an indirect evidence that four point functions are also
zero. Even though we do not have any complete proof of this, it is
natural to expect that this model is a free theory except the
reflection or the two point function due to the Liouville wall. On
the other hand, a similar $c<1$ deformation via a time-like linear
dilaton in bosonic string or type 0 string leads to non-trivial
S-matrix \DK \TLD. This suggests that the $N=2$ non-critical
string has a rather different structure than the $N=0,1$ string
counterparts.

Furthermore, we also study another $N=2$ string model whose matter
part is given by the time-like $N=2$ Liouville theory, which is
defined by a Wick rotation of the ordinary space-like $N=2$
Liouville theory as in the $N=0$ case \ST \SC. This theory will be
closely related to the minimal $\hat{c}<1$ string (or $N=2$ string
on ALE space \OV), which are defined by a combination of a $N=2$
minimal model and the $N=2$ Liouville theory. In this case, we
find that the three particle amplitudes are non-trivial in
general.

The most ambitious goal will be to construct a non-perturbative
description like matrix models\foot{For a discussion on a
holographic dual theory for $N=2$ string on $R^{2,2}$ refer to
\Jev.}. It is natural to believe that this may be possible by
applying an open-closed duality. Indeed, its simple spectrum, i.e.
just a single scalar field, strongly suggests this. Since the rich
spectrum of D-brane boundary states in $N=2$ Liouville theory has
been recently obtained \ES \AR \IPT \Hos , this may not be
difficult (see e.g. \Nopen \GOSD\ for the D-branes in $N=2$ string
on $R^{2,2}$).  We will leave this for a future problem.

This paper is organized as follows. In section 2 we give the
definition of the $N=2$ string whose matter part is the time-like
linear dilaton SCFT. In section 3 we review $N=4$ topological
string formulation which is equivalent to $N=2$ string and we
explain how to apply it to our model. In section 4 we compute the
scattering amplitudes in that theory. In section 5 we study
another $N=2$ string whose matter part is the $N=2$ time-like
Liouville theory.

\newsec{$N=2$ $\hat{c}<1$ String with Time-like Linear Dilaton Matter}

The $N=2$ string has the critical central charge $c=6$ (or
$\hat{c}=2$) and its simplest example is the four dimensional flat
spacetime with two times and two spaces $R^{2,2}$. Even in this
critical case, there is only one physical state (massless scalar
field or self-dual graviton) and thus it looks similar to the
$c=1$ bosonic string or $\hat{c}=1$ type 0 string. Here we
consider the $\hat{c}<1$ deformation by using the time-like linear
dilaton SCFT as a matter part. In section 4 we will analyze
another $N=2$ $\hat{c}<1$ string by adding a time-like Liouville
term.

\subsec{Definition}

In general, $N=2$ string is defined by specifying $N=2$ SCFT which
has the total central charge $\hat{c}=2$ and by gauging the $N=2$
Virasoro algebra generated by $(T,G^{\pm},J)$. Here we assume the
combination of the time-like linear dilaton $N=2$ SCFT
($\hat{c}=1-Q^2$) and the $N=2$ Liouville SCFT ($\hat{c}=1+Q^2$)
as the $\hat{c}=2$ $N=2$ SCFT. We define the world-sheet fields in
the time-like theory by $(X^0,X^1)$ with fermions
$(\psi_0,\bar{\psi_0})$, and those in the space-like Liouville
theory by $(\phi,Y)$ with fermions $(\psi,\bar{\psi})$.  Their
OPEs are \eqn\opescalar{\eqalign{& X^{0}(z)X^{0}(0)\sim \log z,\ \
X^{1}(z)X^{1}(0)\sim \log z,\ \ \psi_0(z)\bar{\psi_0}(0)\sim
-\f{1}{z}, \cr &\phi(z)\phi(0)\sim -\log z,\ \ Y(z)Y(0)\sim -\log
z,\ \ \psi(z)\bar{\psi}(0)\sim \f{1}{z}.}} We can bosonize the
fermions \eqn\bosonize{\psi_0=ie^{iH_0},\
\bar{\psi}_0=ie^{-iH_0},\ {\psi}=e^{iH},\ \bar{\psi}=e^{-iH}.} The
string coupling constant $g_s$ depends on both time and space due
to the null linear dilaton in this background\foot{In this paper
we always set $\al=2$.}
\eqn\acoupling{g_s=e^{\f{Q}{2}(X^0-\phi)}.} Then the $N=2$ energy
stress tensor and their super-partners are \eqn\supertwo{\eqalign{
&T=\f{1}{2}(\de X^0)^2+\f{Q}{2}\de^2 X^0+\f{1}{2}(\de X^1)^2
+\f{1}{2}(\psi_0\de\bar{\psi_0}+\bar{\psi_0}\de\psi_0) \cr &\ \ \
-\f{1}{2}(\de \phi)^2-\f{Q}{2}\de^2 \phi-\f{1}{2}(\de
Y)^2-\f{1}{2}(\psi\de\bar{\psi}+\bar{\psi}\de\psi), \cr &
G^+=-i\psi_0\de(X^0-iX^1)-iQ\de \psi_0+i\psi\de(\phi-iY)+iQ\de
\psi, \cr & G^-=-i\bar{\psi_0}\de(X^0+iX^1)-iQ\de
\bar{\psi}_0+i\bar{\psi}\de(\phi+iY)+iQ\de \bar{\psi}, \cr &
J=-\psi_0\bar{\psi_0}-iQ\de X^1+\psi\bar{\psi}+iQ\de Y .}} In the
closed string theory there are two copies of the generators
\supertwo\ as usual and they are denoted by $(T_{L},G^{\pm}_{
L},J_L)$ and $(T_{R},G^{\pm}_{R},J_R)$ .

Next we introduce the $N=2$ Liouville potential to regulate the
strongly coupled region\foot{In this model, the strongly coupled
region at the time-like infinity is not regulated. However, if we
consider the scattering of a massless particle off the Liouville
wall, the particle will not get into the strongly coupled region.}
\eqn\lppt{\f{\mu}{2\pi} \int dz^2 d\theta^2 e^{-\f{\Phi}{Q}}+
h.c.\ \ ,} where $\Phi$ is the chiral super-field whose lowest
component is given by $\phi+iY$. It is well-known that the $N=2$
Liouville theory is T-dual to the $N=2$ $SL(2,R)_{n}/U(1)$ coset
\HK. The level $n$ is related to the background charge via
$Q=\s{\f{2}{n}}$. In our example we do not have to assume that $n$
is always an integer because we do not couple the $N=2$ Liouville
theory to a $N=2$ minimal model.

Finally we can define a $N=2$ $\hat{c}<1$ string by gauging the
generators $(T_{L,R},G^{\pm}_{ L,R},J_{L,R})$. In the original
model of $N=2$ Liouville theory the radius in $Y$ direction is
given by $Q$. We can also take a $Z_m$ orbifold of the cigar model
$SL(2,R)_{n}/U(1)$ in the circle direction. This is T-dual to the
$N=2$ Liouville theory with the radius $mQ$ \HK. Taking the large
$m$ limit, we can decompactify the $Y$ direction. In this paper we
are mainly interested in the case where $Y$ and $X^1$ are
non-compact\foot{It is straightforward to see that the one-loop
torus partition function in the non-compact case is the same as
the ordinary case $Q=0$ after integrating over the zero modes,
assuming the continuous representation. Thus it is modular
invariant in the $N=2$ string sense.}, though we will also give
results in the compactified case.

\subsec{Physical Vertex Operators and Physical Spacetime
Dimension}

Let us first consider the case where the radius of $Y$ is
infinite. The basic physical state in the $N=2$ $\hat{c}<1$ models
is a single massless scalar field as in the ordinary $N=2$ string
on $R^{2,2}$. In the $N=2$ string, since the $U(1)$ R-current is
gauged, we have only to consider the $(NS,NS)$ sector. Other spin
structures are equivalent.

The vertex operator in the $(-1,-1,-1,-1)$ picture can be written
as
\eqn\vertextal{V=e^{-\phi_{1L}-\phi_{2L}-\phi_{1R}-\phi_{2R}}\cdot
\exp\left[\left(\f{Q}{2}+iE_0\right)X^0
+iE_1X^1+\left(-\f{Q}{2}+ik_2\right)\phi+ik_3Y\right],} where
$\phi_{1}$ and $\phi_{2}$ are the bosonized superconformal ghosts
in $N=2$ string. The physical conditions that the conformal
dimension $\Delta$ of $V$ is one and its R-charge $q$ is zero,
lead to the constraints \eqn\onshellconf{E_0=\pm k_2, \ \ \ \ \ \
\ E_1=-k_3.} The second condition does not have any bosonic or
$N=1$ string counterpart, while the other one comes from the
on-shell condition. This extra condition reduces the original
$2+2$ dimensional spacetime to the three dimensional one which has
one time, one space and one null direction\foot{We can also
consider another
 model by taking the T-duality in the $Y$ direction.
 This is so called $\ap$ type in \COY (see also \Hull), while the
 ordinary one is called $\beta$ type.
In the $\ap$ type case, the constraints are given by the
asymmetric ones $E_0^{L,R}=\pm k^{L,R}_2, \ \ \ \ E^L_1=-k^L_3,\ \
\ E^R_1=k^R_3$. If we assume $Y$ and $X^1$ are non-compact, we
find that the physical spacetime is reduced to 1+1 dimensional.
However, in this case it seems difficult to construct a modular
invariant partition function. It is possible that the model is
consistent when compactified. Since this model is related to the
$\beta$ type model via the T-duality, we find their scattering
amplitudes directly from those in this paper. So we only talk
about $\beta$ type model in this paper.}.

This operator \vertextal\ is the only physical state when we
consider the continuous representation in the $N=2$ Liouville
theory as we can see from its partition function. There may be
some discrete states, though we will not discuss them in this
paper. If we compactified $Y$ and $X^1$, then we need to consider
winding modes also. The physical state conditions can be analyzed
in the same way.

\newsec{$N=4$ Topological String Description}

Since the $N=2$ SCFT which appears in $N=2$ string always has the
central charge $\hat{c}=2$, its symmetry enhances to the $N=4$
superconformal algebra by adding the $SU(2)$ currents
$J^{++}=e^{\int J}$ and $J^{--}=e^{-\int J}$.

For example, in our model \supertwo, the other generators are
given by \eqn\nfouralg{\eqalign{&
J^{++}=e^{iH_0}e^{iH}e^{iQY-iQX^1}, \cr &
J^{--}=e^{-iH_0}e^{-iH}e^{-iQY+iQX^1}, \cr &
\ti{G}^+=e^{iQY-iQX^1}\left[e^{iH_0}\left(i(\de\phi+i\de Y)+Q\de
H\right) +e^{iH} \left(-(\de X^0+i\de X^1)+iQ\de
H_0\right)\right], \cr &
\ti{G}^-=e^{-iQY+iQX^1}\left[e^{-iH_0}\left(i(\de\phi-i\de Y)-Q\de
H\right) +e^{-iH} \left(-(\de X^0-i\de X^1)-iQ\de
H_0\right)\right].}}

By employing this $N=4$ symmetry, we can define $N=4$ topological
string theory and this is known to be equivalent to $N=2$ string
\BV\ in general. Since in our example it turns out that the
computations are easier and more systematic in the $N=4$
topological string description, we will compute the scattering
amplitudes in this formulation later.

\subsec{$N=4$ Topological String and Tree Level Scattering
Amplitudes}

Here we review the definition of $N=4$ topological string and the
computations of on-shell scattering amplitudes. Its topological
twist can be given by $T\to T+\f{1}{2}\de J$ as usual \EY \BV.
After the twist the operators $G^+$ and $\ti{G}^+$ have the
conformal dimension $\Delta=1$, while $G^-$ and $\ti{G}^-$ have
$\Delta=2$. Since the former ones satisfy
$(G^+_0)^2=(\ti{G}^+_0)^2=\{G^+_0,\ti{G}^+_0\}=0$, they
 behave like BRST operators. The physical state is a R-charge $+1$
state that satisfies \eqn\physco{G^+_{0}\Psi=\ti{G}^+_{0}\Psi=0,}
and the equivalence is defined by $\Psi\sim
\Psi+G^+_{0}\ti{G}^+_{0}\chi$.

In this formalism we can define each of the $M$-particle genus $g$
scattering amplitudes \BV. In particular, we are interested in the
most basic one: three point function at tree level in closed
string ($g=0,M=3$). It is defined by \eqn\nfouramot{A_3=\la
|\widehat{\ti{G}}^+_{0}|^2V \cdot |\widehat{\ti{G}}^+_{0}|^2V\cdot
V \lb,} where the square means the left-moving and right-moving
sector contributions. The $\Delta=q=0$  operator\foot{This
operator $V$ is interpreted as the $(-1,-1,-1,-1)$ picture vertex
operator in the $N=2$ string as will be clear from the later
arguments.} $V$ is related to the physical state\foot{In many
examples we can find a corresponding $V$ operator when $\Psi$ is
given \BV. Indeed it is easy to check that the vertex \vertextal\
satisfies $G^+_{0}\ti{G}^+_{0}V=\ti{G}^+_{0}G^+_{0}V=0$ by using
the on-shell conditions in our examples.} $\Psi$ in $N=4$
topological string via \eqn\pjjj{\Psi= |\widehat{\ti{G}^+_{0}}|^2
V,} where we defined $\widehat{\ti{G}^+}$ by \eqn\twistor{
\widehat{\ti{G}^+}=u_1 \ti{G}^++u_2 G^+.} The twistor parameters
$(u_1,u_2)$ satisfy $|u_1|^2+|u_2|^2=1$ and correspond to the
$SU(1,1)$ Lorentz transformation. The parameters for the
left-moving sector or right-moving sector are denoted by
$(u^L_{1},u^L_{2})$ or $(u^R_{1},u^R_{2})$.

It is easy to confirm that the three point function \nfouramot\ in
the $N=4$ topological string indeed agrees with that in $N=2$
string. The $N=2$ string amplitudes with the zero instanton number
is written as $\left\la V\cdot V\cdot
|[G^+_{-1/2},G^{-}_{-1/2}]|^2V \right\lb$. Then we perform the
topological twist and this procedure is equivalent to the spectral
flow by $M-2(1-g)=1$ unit, i.e. the insertion of $e^{\int
J}=J^{++}$. After we move the positions of $J^{++}$ and $G^{-}$
using contour integral expressions and use the OPE
$J^{++}(z)G^{-}(0)\sim z^{-1}\ti{G}^+(0)$, we reproduce the terms
in \nfouramot\ which are proportional to
$u^L_{1}u^L_{2}u^R_{1}u^R_{2}$. In the same way we can confirm
that the other terms correspond to the amplitudes in non-zero
instanton sectors.

Since \nfouramot\ is the topologically twisted expression, we have
to rewrite it in the untwisted NS-sector language in order to
compute it in the CFT expressions. It is given by
\eqn\nfourcft{A_3=\la |\widehat{\ti{G}}^+_{-1/2}V\cdot
\widehat{\ti{G}}^+_{-1/2}V\cdot J^{--}_{-1}V|^2\lb,} after the
spectral flow by $J^{--}=e^{-\int J}$. Then it is obvious that the
total R-charge is zero consistently.

\subsec{An Example of $N=4$ Topological String Amplitudes: Flat
Space $R^{2,2}$}

Let us first consider the $N=2$ string in flat space\foot{Metric
is $g_{1\bar{1}}=1,\ g_{2\bar{2}}=-1$. Here we do not assume any
linear dilaton.} $R^{2,2}$ $(Z^1,\bar{Z^1},Z^2,\bar{Z^2})$ from
the viewpoint of $N=4$ topological string and compute the
amplitudes. In this case we have
\eqn\primayyy{V=e^{i(k_1\bar{Z^1}+\bar{k_1}Z^1+k_2\bar{Z^2}+\bar{k_2}Z^2)}.}
The superconformal generators are \eqn\sugene{\eqalign{&
G^+=\psi^1\de \bar{Z}^1-\psi^2\de \bar{Z}^2,\ \ \
G^-=\bar{\psi}^1\de Z^1-\bar{\psi}^2\de Z^2, \cr &
\ti{G}^+=\psi^1\de Z^2-\psi^2\de Z^1,\ \ \
\ti{G}^-=\bar{\psi}^1\de \bar{Z}^2-\bar{\psi}^2\de \bar{Z}^1. }}
Notice also that $J^-=\psi^1\psi^2$.

Then we obtain \eqn\psiii{\Psi=\left|u_1(k_1\psi_2+k_2\psi_1)+
u_2(\bar{k_1}\psi^1+\bar{k_2}\psi^2)\right|^2
e^{i(k\bar{Z}+\bar{k}Z)}.} The three point function can be
computed as \eqn\three{ A_3=\la
\Psi^{(1)}\Psi^{(2)}|J^-_{-1}|^2V^{(3)}\lb
=c_{12}(u^L_1,u^L_2)\cdot c_{12}(u^R_1,u^R_2),} where we have
defined \eqn\cot{\eqalign{& c_{12}(u_1,u_2) =
(u_1)^2[k^{(1)}_2k^{(2)}_1-k^{(1)}_1k^{(2)}_2]+
(u_2)^2[-\bar{k}^{(1)}_2\bar{k}^{(2)}_1+\bar{k}^{(1)}_1\bar{k}^{(2)}_2]
\cr & \ \
+u_1u_2[\bar{k}^{(1)}_1k^{(2)}_1-\bar{k}^{(2)}_1k^{(1)}_1
-(\bar{k}^{(1)}_2k^{(2)}_2-\bar{k}^{(2)}_2k^{(1)}_2)].}} The
momenta $k^{a}_{1,2}$ represent those of the $a$-th ($a=1,2,3$)
particle.

 Indeed the term proportional to $u^L_1u^R_1u^L_2u^R_2$
in \three\ corresponds to the well-known $N=2$ string amplitude
with the zero instanton number \OVN, while the other terms
proportional to
$(u^L_1)^{1+n_L}(u^L_2)^{1-n_L}(u^L_1)^{1+n_R}(u^L_2)^{1-n_R}$
correspond to the amplitudes in the instanton number $(n_L,n_R)$
sector induced by the $U(1)$ gauge flux on the $N=2$ Riemann
surface.

\newsec{Scattering Amplitudes}

The scattering S-matrix is the most basic quantity which
characterizes string theory models. In this section we discuss the
tree level scattering amplitudes in the $N=2$ $\hat{c}<1$ string
defined in the previous section. The S-matrix for $M_1+M_2$
particle scattering describes the amplitude when we send $M_1$
particles from the large $\phi$ region and observe that $M_2$
particles are reflected back from the Liouville wall\foot{Here we
implicitly neglect the physical interpretation of the extra null
direction. It looks harmless compared with the extra
time-direction in the ordinary $N=2$ string on $R^{2,2}$.} as in
the two dimensional string.

\subsec{Classification of Amplitudes}

As we have shown in section 2.2, the on-shell vertex operators
\vertextal\ can be classified into two types due to the on-shell
conditions \onshellconf : $k_2=E_0$ and $k_2=-E_0$. We call each
of them $(+)$ type (in-coming) or $(-)$ type (out-going),
respectively. When we evaluate the amplitudes in terms of the
correlation functions  in the Liouville theory, it is useful to
employ the quantum numbers $(j,m,\bar{m})$ in the equivalent
$SL(2,R)/U(1)$ coset model. When compactified in $Y$ and $X^1$
direction, the relation between them is given by
\eqn\relabe{ik_2=iE_0=Q(j+1/2), \ \ \ k^L_3=-E^L_1=Qm, \ \ \
k^R_3=-E^R_1=Q\bar{m},} for a $(+)$ particle, while it is
\eqn\relabee{ik_2=-iE_0=Q(j+1/2),  \ \ \ k^L_3=-E^L_1=Qm, \ \ \
k^R_3=-E^R_1=Q\bar{m},} for a $(-)$ particle\foot{ On the other
hand, in the $\ap$ string case, we just have to replace $k^R_3$
with $-k^R_3$ in \relabe\ and \relabee.}.
 By using these quantum numbers instead of the energy and
momenta, we can express the vertex operator for a in-coming (or
out-going) particle by $V^{(+)}_{j,m,\bar{m}}$ (or
$V^{(-)}_{j,m,\bar{m}}$). It is also convenient to define\foot{For
example, we get
$(k^L,\bar{k}^L,E^L,\bar{E}^L)=Q(j+m,j-m,j-m+1,j+m+1)$ for $(+)$
type and for the $(-)$ type
$(k^L,\bar{k}^L,E^L,\bar{E}^L)=Q(j+m,j-m,-j-m,-j+m),$ for the
$(-)$ type.}
 the complex valued momenta\foot{On-shell conditions are given by
$k\bar{k}-E\bar{E}+Qk+QE=k\bar{k}-E\bar{E}+Q\bar{k}+Q\bar{E}=0$.}
\eqn\simppp{k=ik_2+k_3-Q/2,\ \ \ \bar{k}=ik_2-k_3-Q/2,\ \ \
E=iE_0+E_1+Q/2,\ \ \ \bar{E}=iE_0-E_1+Q/2.}

\subsec{Reflection Relation}

As in the bosonic or $N=1$ case, the physical excitations are
reflected back off the Liouville wall in the $N=2$ Liouville
theory. This leads to the following equivalence relation between
\eqn\reflectwg{V^{(+)}_{j,m,\bar{m}}\sim
R(j,m,\bar{m})V^{(-)}_{-j-1,m,\bar{m}},} where we defined
 the reflection coefficient (or two point function) \GK \ES \Hos
 \eqn\reflectcoo{R(j,m,\bar{m})
 =-\mu^{\f{2(2j+1)}{n}}\f{\Gamma(1+j+m)\Gamma(1+j-\bar{m})}
 {\Gamma(-j+m)\Gamma(-j-\bar{m})}
 \f{\Gamma\left(-\f{2j+1}{n}\right)\Gamma(-2j-1)}
 {\Gamma\left(\f{2j+1}{n}\right)\Gamma(2j+1)}.}

\subsec{Three Particle Scattering Amplitudes}

The three point functions are classified into the following four
types \eqn\classty{ (+++),\ \ (++-),\ \ (--+),\ \  (---).} For
each type the $X^0$ momentum conservation can be written as
\eqn\moconxz{j_1+j_2+j_3=-2,\ \ \ j_1+j_2-j_3=-1,\ \ \
j_1+j_2-j_3=0,\ \ \  j_1+j_2+j_3=-1.} Since the $(+)$ and $(-)$
are related to each other by the reflection $j\to -j-1$ as we have
seen in the previous subsection, we just have to compute in one of
the four case. We will argue that all of the three particle
amplitudes are actually vanishing.

Let us first give a short and intuitive
 argument why we find the vanishing amplitudes.
When we consider the $(---)$ amplitude, the energy conservation
condition i.e. the last one in \moconxz : $j_1+j_2+j_3=-1$
coincides with the momentum conservation in the space-like linear
dilaton theory. Then we do not need any insertion of the Liouville
potential to compute it. Thus we expect that the $(---)$ amplitude
is the same as that calculated in the linear dilaton theory. The
three point function in linear dilaton $N=2$ string is generally
proportional to $(c_{12})^2$, where we defined \eqn\kineff{c_{ab}
=E^{(a)}\bar{E}^{(b)}-E^{(b)}\bar{E}^{(a)}-
k^{(a)}\bar{k}^{(b)}+k^{(b)}\bar{k}^{(a)}.} For on-shell particles
they satisfy \eqn\relationmassnd{ c_{ab}=-c_{ba},\ \ \ \ \ \
\sum_{a=1}^{4}c_{ab}=0.} The momentum $k^{(a)}$ and energy
$E^{(a)}$ ($a=1,2,3$) are those for $a$-th particle in the complex
valued notation \simppp. Since $k^{(a)}$ and $\bar{k}^{(a)}$ are
proportional to $E^{(a)}$ and $\bar{E}^{(a)}$ for the $(-)$
particle, $c_{ab}$ is vanishing. Therefore the $(---)$ amplitude
is vanishing\foot{ If we do the same computation for other cases
in the totally linear dilaton model, the amplitude $(++-)$ turns
out to be non-zero in a specific case \AAD. However, this is not
consistent with the reflection relation in the Liouville theory.
We believe at that point the computation becomes singular and the
correct answer is zero.}. The others will also be zero because of
the reflection relation. Notice that we have non-vanishing three
particle scattering amplitudes in $N=0$ (bosonic) and $N=1$
(type0) string. This shows that the non-critical $N=2$ string has
a rather different property than the $N=0,1$ one.

\vskip .1in

More rigorous proof of the vanishing amplitudes including those in
non-zero instanton sectors can be done by the explicit computation
of the amplitudes using the $N=2$ Liouville theory. We will show
the details in the appendix A.2 and here we gave a brief sketch of
the calculations. We employ the equivalent description of the
$N=4$ topological string explained in the previous section.

To make the presentation simple, we assume the $X^1$ direction is
non-compact. Then we only allow the momentum modes in $X^1$ and
$Y$ direction. The results for the compact case including winding
modes can be found in a very similar way and they are also
vanishing.

In this set up, we can write the amplitude \nfourcft\
in the following form \eqn\amothreemid{ A_{3}=
(u^L_1)^2(u^R_1)^2A^{11}+(u^L_2)^2(u^R_2)^2A^{22}
+u^L_1u^L_2u^R_1u^R_2A^{12}. }

For example, the $A^{11}$ can be expressed by the kinematical
factors and the three point functions in $N=2$ Liouville theory.
\eqn\uouof{\eqalign{A^{11}\equiv & \la |\ti{G}^+_{-1/2}V\cdot
\ti{G}^+_{-1/2}V\cdot J^-_{-1}V |^2 \lb \cr & =
 \delta(\sum_{a=1}^3m_a+1)\cdot
\delta(\sum_{a=1}^3\bar{m}_a+1)\cdot\delta(iE^{(a)}_0+\f{Q}{2})
\cdot [-a+b+c-d],}} where $a,b,c$ and $d$ are defined by
\eqn\uouoddd{\eqalign{& a=(\bar{k}^{(1)})^2(\bar{E}^{(2)})^2\cdot
\la
V^{(0,0)}_{j_1,m_1+1,m_1+1}V^{(1,1)}_{j_2,m_2,m_2}V^{(-1,-1)}_{j_3,m_3,m_3}
\lb ,\cr
&b=\bar{k}^{(1)}\bar{k}^{(2)}\bar{E}^{(1)}\bar{E}^{(2)}\cdot \la
V^{(0,1)}_{j_1,m_1+1,m_1}V^{(1,0)}_{j_2,m_2,m_2+1}V^{(-1,-1)}_{j_3,m_3,m_3}
\lb, \cr &
c=\bar{k}^{(1)}\bar{k}^{(2)}\bar{E}^{(1)}\bar{E}^{(2)}\cdot \la
V^{(1,0)}_{j_1,m_1,m_1+1}V^{(0,1)}_{j_2,m_2+1,m_2}V^{(-1,-1)}_{j_3,m_3,m_3}
\lb, \cr & d=(\bar{k}^{(2)})^2(\bar{E}^{(1)})^2\cdot \la
V^{(1,1)}_{j_1,m_1,m_1}V^{(0,0)}_{j_2,m_2+1,m_2+1}V^{(-1,-1)}_{j_3,m_3,m_3}
\lb,}} where $V^{(s,\bar{s})}_{j,m,\bar{m}}$ denotes the primary
operator in the $N=2$ Liouville theory defined by
\eqn\primarynt{V^{(s,\bar{s})}_{j,m,\bar{m}}=e^{Qj\phi+iQ(m+s)Y_L
+iQ(\bar{m}+\bar{s})Y_R+isH_L+i\bar{s}H_R}.}

The three point functions in the $N=2$ Liouville theory have
already computed in \GK \Hos\ as reviewed in appendix A.1. After
we substitute them into \uouoddd, we can find that $A^{11}$ is
actually zero for all four cases \classty. In the $(---)$ case
this is very easy to see because all of the Liouville three point
functions in \uouoddd\ are the same. It is indeed proportional to
the $(c_{12})^2$ as expected from the previous linear dilaton
calculation. In the same way we can confirm that $A^{12}$ and
$A^{22}$ are vanishing as we have shown in the appendix A.2.

\subsec{Discussions on Four Particle Scattering}

Even though the explicit four point function in $N=2$ Liouville
theory is not known, we can discuss the four particle scattering
from the analysis of the linear dilaton theory. Motivated by this,
let us compute the on-shell four particle scattering in the linear
dilaton theory i.e. $\mu=0$. We define the Mandelstam valuables
($a=1,2,3,4$ represents the label of the four particles)
\eqn\mandel{\eqalign{s&=-(E^{(1)}\bar{E}^{(3)}+E^{(3)}\bar{E}^{(1)})
+k^{(1)}\bar{k}^{(3)}+k^{(3)}\bar{k}^{(1)}, \cr
t&=-(E^{(2)}\bar{E}^{(3)}+E^{(3)}\bar{E}^{(2)})
+k^{(2)}\bar{k}^{(3)}+k^{(3)}\bar{k}^{(2)}, \cr
u&=-(E^{(4)}\bar{E}^{(3)}+E^{(3)}\bar{E}^{(4)})
+k^{(4)}\bar{k}^{(3)}+k^{(3)}\bar{k}^{(4)}.}} They satisfy the
relations $s+t+u=0$ for on-shell particles.

Then the S-matrix of the four particles is given by
\eqn\ampfourldld{A_{4}=F^2\cdot
\f{\Gamma(1-s/2)\Gamma(1-t/2)\Gamma(1-u/2)}
{\Gamma(s/2)\Gamma(t/2)\Gamma(u/2)},} where $F$ is defined by
\eqn\valuefff{F=1-\f{c_{24}c_{13}}{su}-\f{c_{23}c_{14}}{tu}.} This
resulting form is the same as the well-known result\foot{ Notice
that we need the exchange $1\lr 4$ due to the different
convention.} in flat space \OVN, though the values of momenta
$E,k$ are different from that one. We expect that the $(----)$
amplitude in the $N=2$ Liouville theory should coincide with the
one in the linear dilaton theory as we confirmed in the previous
example. In this case, $c_{ab}$ and $(s,t,u)$ are all vanishing.
By treating them as infinitesimals of the same order, we can find
that the amplitude is zero. Then all other four particle
amplitudes will be zero due to the reflection relation. This may
be analogous to the fact that all $M\geq 4$ particle scattering
amplitudes in flat space $R^{2,2}$ are zero at any genus \BV. This
result strongly suggests that the $N=2$ $\hat{c}<1$ string with
the time-like linear dilaton is actually a free theory\foot{This
is true for both $\ap$ and $\beta$ type model. In the $\ap$ string
model this may not be so surprising because we know that this
theory is free on $R^{2,2}$ without linear dilaton as shown in
\Hull \GOS.} except the reflection (or two point function) at the
Liouville wall. It would be interesting to study this further.

We would also like to note that the naive $Q=0$ limit of our model
is not equivalent to the familiar $N=2$ string on $R^{2,2}$. The
latter has a different dimension and non-zero three particle
scattering amplitudes.

\newsec{Scattering Amplitudes in $\hat{c}<1$ String with Time-like
Liouville Matter}

We can define another $N=2$ $\hat{c}<1$ string by combining a
time-like $N=2$ Liouville theory with the standard (space-like)
$N=2$ Liouville Theory. This time-like $N=2$ Liouville theory can
be defined by the Wick rotation of the space-like one as was done
for the bosonic string case \ST \SC \HT. This $\hat{c}<1$ string
is closely related to the $N=2$ minimal $\hat{c}<1$ string after
the $Z_n$ orbifold. The latter is defined by combining the $N=2$
minimal models and the $N=2$ Liouville theory, which is equivalent
to the $N=2$ string on ALE spaces \OV \AH \KPS. Since the $n$-th
$N=2$ minimal models or equally $N=2$ $SU(2)/U(1)$ coset at level
$n$, is also regarded as the negative level $n\to -n$ continuation
of $N=2$ $SL(2,R)/U(1)$ coset or $N=2$ Liouville theory. Thus we
expect the correlation functions in our model are essentially the
same as those in the minimal $N=2$ string at tree level, though
the number of physical states is restricted to be finite only in
the minimal model case.

\subsec{Time-like $N=2$ Liouville Theory}

We can define the time-like $N=2$ Liouville theory $(X^0,X^1)$
with the background charge $Q_0$ via the Wick-rotation of the
ordinary $N=2$ Liouville theory $(Y,\phi)$ with the background
charge $Q_2$ \eqn\wick{X_0=-i\phi,\ \ X_1=-iY,\ \ H_0=H,\ \
Q_2=iQ_0.} In this case, the momenta are rotated as follows
\eqn\rotatem{E_0=ik_2=iQ_0(j+\f{1}{2}),\ \ \ E^L_1=k^L_3=-Q_0m,\ \
\  E^R_1=k^R_3=-Q_0\bar{m}} where $(j,m,\bar{m})$ are the usual
quantum numbers of the $SL(2,R)_{n+2}$ WZW model
($Q_2=\s{\f{2}{n}}$). In the actual computations of correlation
functions, we set $Q_0=Q$ and regard the time-like vertex
operators as a $SL(2,R)_{2-n}$ primary via the rule
\eqn\rotatem{iE_0=-Q(j+\f{1}{2}),\ \ \ E_1=-Qm.} Notice the flip
of the sign $n\to -n$. The quantum number $s$ is the same as the
space-like theory.

The $(+)$ type $(E_0=k_2)$ physical vertex in this theory is given
by \eqn\vertpu{V^{(+)}_j=\ti{V}_{-j_-1,m,\bar{m}}\otimes
V_{j,m,\bar{m}},} and the $(-)$ type $(E_0=-k_2)$ one is
\eqn\vertpue{V^{(-)}_j=\ti{V}_{j,m,\bar{m}}\otimes
V_{j,m,\bar{m}}.} Here $\ti{V}$ denotes the primary in the
time-like Liouville theory, which can be treated as the ordinary
$N=2$ Liouville theory with the imaginary background charge $iQ$.
$V$ is the primary in the ordinary $N=2$ Liouville theory with the
background charge $Q$. If we are interested in the $N=2$ string on
ALE space (or $\f{SU(2)_{n-2}}{U(1)}\times
\f{SL(2,R)_{n+2}}{U(1)}$), the values of $(j,m)$ become
half-integers and follow constraints e.g. $0\leq j\leq n/2$ as
usual.

\subsec{Three Particle Scattering Amplitudes}

Now we can compute the three particle scattering amplitudes in the
$N=4$ topological string formulation. We present the detailed
calculations in the appendix A.3. Here we just write down the
final result. We assume the compactification of $X^1$ and $Y$ is
arbitrary\foot{Here we present results in the $\beta$ type model
again, but those in $\ap$ type model can be obtained easily via
T-duality.}. The result of the $(---)$ amplitude can be written as
\eqn\finalsaf{A_{3(---)}=A_{\Delta s=0}+A_{\Delta s=\pm 1}.} The
first term comes from the fermion number conserving part of the
three point function (i.e. the first term in (A.1)) and the second
one from the fermion number violating part\foot{Notice that in the
time-like linear dilaton case we only had the fermion number
conserving part since the fermion number in the time-like CFT is
conserved.} (the second and third term in (A.1)).

The latter is explicitly given by \eqn\finalsa{\eqalign{A_{\Delta
s=\pm 1}&=Q^4 (u^L_1-iu^L_2)^2(u^R_1-iu^R_2)^2\cdot
\left[\delta^2\left(\sum_a m_a+1-\f{n}{2}\right)+
\delta^2\left(\sum_a m_a+1+\f{n}{2}\right)\right] \cr &\ \ \ \cdot
\ti{D}_{+}(j_a)\cdot D_{+}(j_a)\cdot
\prod_{a=1}^{3}\f{\Gamma(j_a-m_a+1)\Gamma(j_a+m_a+1)}
{\Gamma(-j_a+\bar{m}_a)\Gamma(-j_a-\bar{m}_a)},}} where
$\ti{D}_{+}(j_a)$ is the wick-rotated (or time-like) version\foot{
This Wick-rotation procedure should be careful and non-trivial as
in \SC \HT.} of the function $D_{+}(j_a)$. The function
$D_{+}(j_a)$ was first defined in  \Hos\ and is also reviewed in
the appendix A.1 of the present paper. Notice that in this case we
have left-right asymmetric terms in addition to the previous
expression \amothreemid . To get amplitudes for the $(+)$
particle, we just have to replace
$\f{\Gamma(j_a-m_a+1)}{\Gamma(-j_a+\bar{m}_a)}$ with
$\f{\Gamma(-j_a-m_a)}{\Gamma(j_a+\bar{m}_a+1)}$ , and
$\ti{D}_{+}(j_a)$ with $\ti{D}_{+}(-j_a-1)$ in \finalsa.

The former term $A_{\Delta s=0}$ in general can be written in
terms of the functions $D(j_a)$ and $F(j_a,m_a,\bar{m}_a)$ as we
computed in the appendix A.3. Actually, it is vanishing\foot{In
more general, we can show $A_{\Delta s=0}=0$ when
$F(j_a,m_a,\bar{m}_a)$ is factorized as
$F(j_a,m_a,\bar{m}_a)=f(j_a,m_a)g(j_a,\bar{m}_a)$ for some
functions $f$ and $g$.} in the following important cases: (i)
$j_a=m_a=\bar{m}_a$ for one of $a=1,2,3$, (ii) either of the four
conditions in \moconxz\ is satisfied.

Finally, let us evaluate \finalsa\ in the special case of
$j_a=m_a=\bar{m}_a$. This correspond to the chiral primary
states\foot{This computation looks very similar to that of three
point functions in the $N=2$ topologically twisted $SL(2,R)/U(1)$
model in \TMV. Indeed, we can derive the same result in \TMV\
using a similar analysis in $N=2$ Liouville theory. Notice that
its result is somewhat different from the $N=4$ case as their
definitions are not the same.}. In this case we can obtain the
explicit formula of $\ti{D}_{+}(j_a)$ and $D_{+}(j_a)$.

The amplitudes are vanishing except $(i)$ $\sum_a
j_a+1-\f{n}{2}=0$ or $(ii)$ $\sum_a j_a+1+\f{n}{2}=0$ is
satisfied. In these two cases we obtain\foot{Naively we may just
regard these amplitudes are vanishing. This is consistent with its
original definition in the $N=4$ topological string because the
$\hat{\ti{G}}_+$ annihilates the chiral primary operator $j=m$.
However, as we will see shortly the $A_{3(+++)}$ is not vanishing.
and this suggests us that $A_{3(---)}$ should not be regarded as
just a zero.} $A^{(i)}_{3(---)}=\ap^{(-)}\cdot \mu$ and
$A^{(ii)}_{3(---)}=-\ap^{(-)}\cdot \mu^{-1}$, where $\ap^{(-)}$
represents a double zero $\sim \Gamma(0)^{-2}$ due to the
Gamma-function in \finalsa. On the other hand, another one
$A_{3(+++)}$ with the opposite chiralities is given by in each
cases \eqn\achione{\eqalign{A^{(i)}_{3(+++)}&= \ap^{(+)}
\cdot\mu\cdot \prod_{a=1}^{3}\gamma\left(1-\f{2j_a+1}{n}\right),
\cr A^{(ii)}_{3(+++)}&= \ap^{(+)} \cdot\mu^{-1}\cdot
\prod_{a=1}^{3}\gamma\left(1-\f{2j_a+1}{n}\right),}} where
$\ap^{(+)}$ is a divergent constant ($\sim \Upsilon(0)^{-1}$). We
also defined $\gamma(x)\equiv \Gamma(x)/\Gamma(1-x)$.

It will be exciting to note that the amplitude $A^{(ii)}_{3(+++)}$
with the momentum conservation $\sum_a j_a+1+\f{n}{2}=0$ in
\achione\ agrees with those in the $(1,n)$ non-critical bosonic
string. This suggests that the chiral primary sector in the
minimal $N=2$ string is essentially equivalent to the $(1,n)$
string. A similar observation can be obtained from the ground-ring
structure computed recently in \KPS. Indeed it includes a
$Z_{n-1}$ subring, which is the same as the ground ring in $(1,n)$
string. Also it is known that the $(1,n)$ string is described by
so called ADE matrix model \ADE. The ADE series of this model seem
to match with the ADE classification of ALE spaces. It would be
interesting to explore this relation further.

\vskip 0.8in

\centerline{\bf Acknowledgments}

I am very grateful to G. Giribet, D. Jatkar, S. Minwalla, L. Motl,
Y. Nakayama, H. Ooguri, Y. Oz, S. Panda, N. Saulina, C. Vafa for
useful discussions and correspondences and especially to K.
Hosomichi for helpful comments. This work was supported in part by
DOE grant DE-FG02-91ER40654.

\vskip 3in

\appendix{A}{Details of Computations of Three Particle Amplitudes}

\subsec{Three Point Functions in $N=2$ Liouville Theory}

The three point functions in $N=2$ Liouville theory are given
by\foot{Here we omit the world-sheet coordinate dependence and the
cocycle factors. It turns out that they are irrelevant for our
computations.}\Hos\ (see also \GK \Gir) \eqn\correnl{\eqalign{&\la
V^{(s_1,\bar{s_1})}_{j_1,m_1,\bar{m}_1}V^{(s_2,\bar{s_2})}_{j_2,m_2,\bar{m}_2}
V^{(s_3,\bar{s_3})}_{j_3,m_3,\bar{m}_3} \lb \cr
&=\delta^2(\sum_{a=1}^3m_a)\cdot \delta^2(\sum_{a=1}^3s_a)\cdot
D(j_a)\cdot F(j_a,m_a,\bar{m}_a) \cr & \ +
\delta^2(\sum_{a=1}^3m_a-1-\f{n}{2})\cdot
\delta^2(\sum_{a=1}^3s_a+1)\cdot D_{-}(j_a)\cdot
F_{-}(j_a,m_a,\bar{m}_a) \cr & \ +
\delta^2(\sum_{a=1}^3m_a+1+\f{n}{2})\cdot\delta^2(\sum_{a=1}^3s_a-1)\cdot
D_{+}(j_a)\cdot F_{+}(j_a,m_a,\bar{m}_a).}}

The functions $D(j_a)$, $D_{\pm}(j_a)$ and
$F_{\pm}(j_a,m_a,\bar{m}_a)$ are given by \eqn\degfffu{\eqalign{&
D(j_a)=\f{(\nu
b^{-2b^2})^{j_1+j_2+j_3+1}\Upsilon(0)'\Upsilon(b(2j_1+1))
\Upsilon(b(2j_2+1))\Upsilon(b(2j_3+1))}{\s{2}b^2\Upsilon(b^{-1}-b(j_{1+2+3}+1))
\Upsilon(b^{-1}+bj_{1-2-3})\Upsilon(b^{-1}+bj_{2-1+3})
\Upsilon(b^{-1}+bj_{3-1-2})}, \cr & D_{\pm}(j_a)=\f{(\nu
b^{2-2b^2})^{j_1+j_2+j_3+1}\Upsilon(0)'\Upsilon(b(2j_1+1))
\Upsilon(b(2j_2+1))\Upsilon(b(2j_3+1))}{\s{2}b^{1+n}\Upsilon(\f{1}{2b}-b(j_{1+2+
3}+1))
\Upsilon(\f{1}{2b}+bj_{1-2-3})\Upsilon(\f{1}{2b}+bj_{2-1+3})
\Upsilon(\f{1}{2b}+bj_{3-1-2})}, \cr &
F_{\pm}(j_a,m_a,\bar{m}_a)=(-1)^{m_2-\bar{m}_2}\prod_{a=1}^3\f{\Gamma(1+j_a\pm
m_a)}{\Gamma(-j_a\mp \bar{m}_1)},}} where we have defined
$b=1/\s{n}$ and $\nu=\mu^{2/n}$; we also used the notation like
$j_{3-1-2}\equiv j_3-j_1-j_2$. The function $\Upsilon(x)$ is
familiar one already in the bosonic Liouville theory \COL. It
satisfies the relations
$\Upsilon(x+b)=\gamma(bx)b^{1-2bx}\Upsilon(x)$ and
$\Upsilon(x+1/b)=\gamma(x/b)b^{2x/b-1}\Upsilon(x)$
($\gamma(x)\equiv \Gamma(x)/\Gamma(1-x)$). Another function
$F(j_a,m_a,\bar{m}_a)$ is defined by the following integral
\eqn\fdefep{\eqalign{& F(j_a,m_a,\bar{m}_a)=\pi^{-2}\int dz^2 dw^2
z^{j_1-m_1}\bar{z}^{j_1-\bar{m}_1}(1-z)^{j_2-m_2}(1-\bar{z})^{j_2-\bar{m}_2}\cr
&\ \  \ \ \cdot
w^{j_1+m_1}\bar{w}^{j_1+\bar{m}_1}(1-w)^{j_2+m_2}(1-\bar{w})^{j_2+\bar{m}_2}
|z-w|^{-4-2(j_1+j_2+j_3)}.}}

\vskip 2in

{\bf Explicit Evaluations of $D(j_a)$ and $F(j_a,m_a,\bar{m}_a)$}

To perform actual calculations it is important to rewrite $D(j_a)$
and $F(j_a,m_a,\bar{m}_a)$ in terms of more basic functions. This
is possible when the energy conservation (i.e. \moconxz ) of the
time-like linear dilaton theory is satisfied. The function
$D(j_a)$ for each case is given by \eqn\dfunce{\eqalign{&
j_1+j_2+j_3=-2:\ \ D(j_a)=2^{-1/2}\nu^{-1}n^{-3}\cdot
\f{\Upsilon'(0)}{\Upsilon(Q)} \cdot \prod_{a=1}^{3}\left[\gamma
(-2j_a)\cdot\gamma (-(2j_a+1)/n)\right], \cr & j_1+j_2-j_3=-1:\ \
D(j_a)=2^{-1/2}\nu^{2j_3}n^{2} \cdot \f{\Upsilon'(0)}{\Upsilon(Q)}
\cdot \gamma \left(1-\f{2j_1+1}{n}\right)\cdot \gamma
\left(1-\f{2j_2+1}{n}\right), \cr & j_1+j_2-j_3=0:\ \
D(j_a)=2^{-1/2}\nu^{2j_3+1}n^{3/2}  \cdot \f{\Upsilon'(0)}
{\Upsilon(b^{-1})} \cdot \gamma \left(1-\f{2j_3+1}{n}\right), \cr
& j_1+j_2+j_3=-1:\ \  D(j_a)=2^{-1/2}n^{1/2}  \cdot
\f{\Upsilon'(0)}{\Upsilon(b^{-1})} \cdot
\prod_{a=1}^{3}\gamma(-2j_a).}} Notice the first two are divergent
because $\Upsilon(Q)=0$. Notice also
$\Upsilon'(0)=\Upsilon(b^{-1})=\Upsilon(b)$.

The function $F(j_a,m_a,\bar{m}_a)$ can also be explicitly
evaluated by employing the formula \eqn\formulaintg{\int dx^2
|x|^{2a}x^n|1-x|^{2b}(1-x)^m=\pi\cdot
\f{\Gamma(a+n+1)\Gamma(b+m+1)\Gamma(-a-b-1)}{\Gamma(-a)\Gamma(-b)\Gamma(a+b+m+n+
2)},} and the reflection relation. The results are given by
\eqn\forreflrm{\eqalign{j_1+j_2+j_3=-2:\ \
F(j_a,m_a,\bar{m}_a)&=\prod_{a=1}^{3}
\f{\Gamma(j_a-m_a+1)\Gamma(j_a+m_a+1)} {
\Gamma(-j_a-\bar{m}_a)\Gamma(-j_a+\bar{m}_a)}, \cr
j_1+j_2-j_3=-1:\ \ F(j_a,m_a,\bar{m}_a)&=(-1)^{m_3-\bar{m}_3}\cdot
\gamma(-2j_1-1)\cdot \gamma(-2j_2-1) \cr &\ \ \cdot
\prod_{a=1}^{2} \f{\Gamma(j_a-m_a+1)\Gamma(j_a+m_a+1)} {
\Gamma(-j_a-\bar{m}_a)\Gamma(-j_a+\bar{m}_a)}, \cr j_1+j_2-j_3=0:\
\ F(j_a,m_a,\bar{m}_a)&=(-1)^{m_1-\bar{m}_1}\cdot \Gamma(0)\cdot
\gamma(-2j_1-1) \cr &\ \ \cdot
\f{\Gamma(j_1-m_1+1)\Gamma(j_1+m_1+1)}{
\Gamma(-j_1-\bar{m}_1)\Gamma(-j_1+\bar{m}_1)}, \cr
j_1+j_2+j_3=-1:\ \ F(j_a,m_a,\bar{m}_a)&=\Gamma(0)\cdot
\prod_{a=1}^3\gamma(2j_a+1).}}

It is also useful to compute this factor in $j_2=m_2=\bar{m}_2$
case. The result is given by \eqn\forreflcp{\eqalign{
F(j_a,m_a,\bar{m}_a)|_{j_2=m_2=\bar{m}_2}
&=(-1)^{m_3-\bar{m}_3}\f{\Gamma(j_1-m_1+1)\Gamma(j_3-m_3+1)}
{\Gamma(-j_1+\bar{m}_1)\Gamma(-j_3+\bar{m}_3)}\cdot
\gamma(2j_2+1)\cr &\cdot \gamma(j_1-j_2-j_3)\cdot
\gamma(-j_1-j_2-j_3-1) \cdot \gamma(-j_1-j_2+j_3) .}}

\subsec{Three Particle Scatterings in $N=2$ $\hat{c}<1$
string with Time-like Linear Dilaton}

Now we would like to evaluate the three particle scattering
amplitudes \amothreemid\ by applying the previous formula in $N=2$
Liouville theory. To compute the amplitude we need the following
expressions \eqn\exrrgt{\eqalign{& G^+_{-1/2}V= \left(-ik e^{iH}
+E e^{iH_0}\right)\cdot e^{(\f{Q}{2}+iE_0)X^0
+iE_1X^1+(-\f{Q}{2}+ik_2)\phi+ik_3Y}, \cr &
\ti{G}^+_{-1/2}V=\left(-i\bar{k}e^{iH_0} -\bar{E}e^{iH}\right)
\cdot e^{(\f{Q}{2}+iE_0)X^0
+i(E_1-Q)X^1+(-\f{Q}{2}+ik_2)\phi+i(k_3+Q)Y}, \cr &
J^{-}_{-1}V=e^{-iH_0}e^{-iH}e^{(\f{Q}{2}+iE_0)X^0
+i(E_1+Q)X^1+(-\f{Q}{2}+ik_2)\phi+i(k_3-Q)Y}.}}

Then it is straightforward to see that the $A^{11}$ is given by
\uouof\ and \uouoddd. Below we also present other amplitudes
$A^{22}$ and $A^{12}$ explicitly. The amplitude $A^{22}$ can be
computed as follows (again we show results in $\beta$ string, but
we can find the results in $\ap$ string easily via T-duality)
\eqn\uouo{\eqalign{A^{22}\equiv & \la |G^+_{-1/2}V\cdot
G^+_{-1/2}V\cdot J^-_{-1}V|^2\lb \cr & = \delta(\sum_{a=1}^3 m_a
-1)\cdot\delta(\sum_{a=1}^3 \bar{m}_a -1)\cdot \delta(\sum_{a=1}^3
iE^{(a)}_0+\f{Q}{2})\cdot [-a'+b'+c'-d'],}} where $a',b',c',d'$
are given by \eqn\abcdf{\eqalign{& a'=(E^{(1)})^2(k^{(2)})^2 \cdot
\la
V^{(0,0)}_{j_1,m_1,m_1}V^{(1,1)}_{j_2,m_2-1,m_2-1}V^{(-1,-1)}_{j_3,m_3,m_3}
\lb ,\cr & b'=E^{(1)}E^{(2)}k^{(1)}k^{(2)}\cdot \la
V^{(0,1)}_{j_1,m_1,m_1-1}V^{(1,0)}_{j_2,m_2-1,m_2}V^{(-1,-1)}_{j_3,m_3,m_3}
\lb \cr & c'=E^{(1)}E^{(2)}k^{(1)}k^{(2)}\cdot \la
V^{(1,0)}_{j_1,m_1-1,m_1}V^{(0,1)}_{j_2,m_2,m_2-1}V^{(-1,-1)}_{j_3,m_3,m_3}
\lb \cr & d'=(E^{(2)})^2(k^{(1)})^2\cdot \la
V^{(1,1)}_{j_1,m_1-1,m_1-1}V^{(0,0)}_{j_2,m_2,m_2}V^{(-1,-1)}_{j_3,m_3,m_3}
\lb .}}

 Also $A^{12}$ is given
by the more complicated expression\foot{Notice that if all the
correlators in the space-like Liouville theory are the same, the
result is proportional to $(c_{12})^2$.} \eqn\morecomp{\eqalign{
A^{12}&\equiv  \la |(G^+_{-1/2}V\cdot \ti{G}^+_{-1/2}V\cdot
J^-_{-1}V+\ti{G}^+_{-1/2}V\cdot G^+_{-1/2}V\cdot J^-_{-1}V)|^2 \lb
\cr & = \delta(\sum_{a=1}^3 m_a)\cdot\delta(\sum_{a=1}^3
\bar{m}_a) \cdot \delta(\sum_{a=1}^3 iE^{(a)}_0+\f{Q}{2}) \cr &
\cdot \Bigl[\left(k^{(1)}\bar{k}^{(2)}\right)^2\cdot \la
V^{(1,1)}_{j_1,m_1-1,m_1-1}V^{(0,0)}_{j_2,m_2+1,m_2+1}V^{(-1,-1)}_{j_3,m_3,m_3}
\lb \cr &- k^{(1)}\bar{k}^{(2)}E^{(1)}\bar{E}^{(2)}\cdot \la
V^{(1,0)}_{j_1,m_1-1,m_1}V^{(0,1)}_{j_2,m_2+1,m_2}V^{(-1,-1)}_{j_3,m_3,m_3}
\lb \cr &-k^{(1)}\bar{k}^{(2)}E^{(1)}\bar{E}^{(2)}\cdot \la
V^{(0,1)}_{j_1,m_1,m_1-1}V^{(1,0)}_{j_2,m_2+1,m_2}V^{(-1,-1)}_{j_3,m_3,m_3}
\lb \cr & +\left(E^{(1)}\bar{E}^{(2)}\right)^2\cdot \la
V^{(0,0)}_{j_1,m_1,m_1}V^{(1,1)}_{j_2,m_2,m_2}V^{(-1,-1)}_{j_3,m_3,m_3}
\lb \cr
%%%%
&+\left(k^{(2)}\bar{k}^{(1)}\right)^2\cdot \la
V^{(0,0)}_{j_1,m_1+1,m_1+1}V^{(1,1)}_{j_2,m_2-1,m_2-1}V^{(-1,-1)}_{j_3,m_3,m_3}
\lb \cr &- k^{(2)}\bar{k}^{(1)}E^{(2)}\bar{E}^{(1)}\cdot \la
V^{(0,1)}_{j_1,m_1+1,m_1}V^{(1,0)}_{j_2,m_2-1,m_2}V^{(-1,-1)}_{j_3,m_3,m_3}
\lb \cr &-k^{(2)}\bar{k}^{(1)}E^{(2)}\bar{E}^{(1)}\cdot \la
V^{(1,0)}_{j_1,m_1,m_1+1}V^{(0,1)}_{j_2,m_2,m_2-1}V^{(-1,-1)}_{j_3,m_3,m_3}
\lb \cr & +\left(E^{(2)}\bar{E}^{(1)}\right)^2\cdot \la
V^{(1,1)}_{j_1,m_1,m_1}V^{(0,0)}_{j_2,m_2,m_2}V^{(-1,-1)}_{j_3,m_3,m_3}
\lb \cr
%%%%
&+k^{(1)}\bar{k}^{(2)}\bar{E}^{(1)}E^{(2)}\cdot \la
V^{(1,1)}_{j_1,m_1-1,m_1}V^{(0,0)}_{j_2,m_2+1,m_2}V^{(-1,-1)}_{j_3,m_3,m_3}
\lb \cr & -k^{(1)}\bar{k}^{(1)}k^{(2)}\bar{k}^{(2)}\cdot \la
V^{(1,0)}_{j_1,m_1-1,m_1+1}V^{(0,1)}_{j_2,m_2+1,m_2-1}V^{(-1,-1)}_{j_3,m_3,m_3}
\lb \cr & -E^{(1)}\bar{E}^{(1)}\bar{E}^{(2)}E^{(2)}\cdot \la
V^{(0,1)}_{j_1,m_1,m_1}V^{(1,0)}_{j_2,m_2,m_2}V^{(-1,-1)}_{j_3,m_3,m_3}
\lb \cr & +E^{(1)}\bar{E}^{(2)}k^{(2)}\bar{k}^{(1)}\cdot \la
V^{(0,0)}_{j_1,m_1,m_1+1}V^{(1,1)}_{j_2,m_2,m_2-1}V^{(-1,-1)}_{j_3,m_3,m_3}
\lb \cr
%%%%%%%
&+k^{(1)}\bar{k}^{(2)}\bar{E}^{(1)}E^{(2)}\cdot \la
V^{(1,1)}_{j_1,m_1,m_1-1}V^{(0,0)}_{j_2,m_2,m_2+1}V^{(-1,-1)}_{j_3,m_3,m_3}
\lb \cr & -k^{(1)}\bar{k}^{(1)}k^{(2)}\bar{k}^{(2)}\cdot \la
V^{(0,1)}_{j_1,m_1+1,m_1-1}V^{(1,0)}_{j_2,m_2-1,m_2+1}V^{(-1,-1)}_{j_3,m_3,m_3}
\lb \cr & -E^{(1)}\bar{E}^{(1)}\bar{E}^{(2)}E^{(2)}\cdot \la
V^{(1,0)}_{j_1,m_1,m_1}V^{(0,1)}_{j_2,m_2,m_2}V^{(-1,-1)}_{j_3,m_3,m_3}
\lb \cr & +E^{(1)}\bar{E}^{(2)}k^{(2)}\bar{k}^{(1)}\cdot \la
V^{(0,0)}_{j_1,m_1+1,m_1}V^{(1,1)}_{j_2,m_2-1,m_2}V^{(-1,-1)}_{j_3,m_3,m_3}
\lb \Bigr].}}

\vskip .2in

{\bf Evaluation of the Amplitudes}

Consider the $(---)$ case. Then as we can see from \dfunce\ and
\forreflrm, the three point function in the Liouville theory does
not depend on $m_a(=\bar{m}_a)$. Thus we can factorize the
kinematical factors and find that they are all vanishing because
$(k^{(a)},\bar{k}^{(a)})=-(E^{(a)},\bar{E}^{(a)})$ for (-)
particles. Furthermore, application of the reflection relation or
the direct evaluations show that all other types of amplitudes are
vanishing.

In all four types of on-shell amplitudes the three point function
$\sim D(j_a)F(j_a,m_a,\bar{m}_a)$ in the Liouville theory includes
a common divergence. This is because\foot{ One may wonder a
careful treatment of this divergence in \dfunce\ and in
\forreflrm\ may can be canceled by the zeros in the kinematical
factors leading to non-vanishing finite answers. However, this
does not seem to be the case. Let us regulate by assuming the
extra $\hat{c}=D$ directions. The linear dilaton gradient is now
given by $g_s=e^{\f{q}{2}X_0-\f{Q}{2}\phi}$. Define $q=(1-\ep)Q$
and treat $\ep$ as an infinitesimal constant. Then it is easy to
see that the correlation function for $(---)$ is given by just
replacing $\Gamma(0)$ in \forreflrm\ with $(\ep)^{-1}\cdot
f(j_a,m_a)$, where $f$ is a non-singular function. Then the
kinematical factor is proportional to $\ep^2$, while the Liouville
correlator to $\ep^{-1}$. Thus the total amplitude is of order
$\ep$ and is vanishing.} we do not need any insertion of Liouville
operator in that computation when $j_1+j_2+j_3=-1$.

In the above computations we assumed that $Y$ and $X^1$ are
non-compact directions. In the compact case we have to take
winding modes into account as in \relabee. But this can be done
almost in the same way as before and in the end the three particle
amplitudes are all vanishing.

\subsec{Three Particle Scatterings in $N=2$ $\hat{c}<1$
string with Time-like Liouville}

Finally we present the detailed derivation of the result \finalsa\
in section 5. In this time-like Liouville case, one important
point is that we can allow the violation of the fermion numbers
$s_a$ and $\bar{s}_a$. Indeed as we can see from the explicit
computation below, only such new contributions become non-zero.

To make the expressions simple, we show results only for the
$A^{11}$ amplitude in the $\beta$ type model, which is
proportional to $(u^L_1)^2(u^R_1)^2$. The other part of the
amplitude can be computed in the same way. We take the
compactification in $Y$ and $X^1$ direction arbitrary, including
the non-compact case. It can be expressed in terms of correlation
functions for the space-like and time-like Liouville theory as
follows \eqn\uouofm{\eqalign{A^{11}\equiv & \la
|\ti{G}^+_{-1/2}V\cdot \ti{G}^+_{-1/2}V\cdot J^-_{-1}V |^2 \lb \cr
=& \delta^2\left(\sum_{a=1}^{3}m_a+1\right)\cdot [-a+b+c-d] \cr &
+ \delta^2\left(\sum_{a=1}^{3}m_a+1-\f{n}{2}\right)\cdot [e] +
\delta^2\left(\sum_{a=1}^{3}m_a+1+\f{n}{2}\right)\cdot [f] \ \ ,}}
where $a,b,c,d$ is defined by \eqn\uouodddm{\eqalign{& a=
\bar{k}^{(1)}_{L}\bar{E}^{(2)}_{L}
\bar{k}^{(1)}_{R}\bar{E}^{(2)}_{R} \cdot \la
\ti{V}^{(1,1)}_{\ti{j}_1,m_1,\bar{m}_1}
\ti{V}^{(0,0)}_{\ti{j}_2,m_2+1,\bar{m}_2+ 1}
\ti{V}^{(-1,-1)}_{\ti{j}_3,m_3,\bar{m}_3} \lb \cdot \la
V^{(0,0)}_{j_1,m_1+1,\bar{m}_1+1}V^{(1,1)}_{j_2,m_2,\bar{m}_2}
V^{(-1,-1)}_{j_3,m_3,\bar{m}_3} \lb ,\cr &b=
\bar{k}^{(1)}_{L}\bar{E}^{(2)}_{L}
\bar{k}^{(2)}_{R}\bar{E}^{(1)}_{R} \cdot \la
\ti{V}^{(1,0)}_{\ti{j}_1,m_1,\bar{m}_1+1}
\ti{V}^{(0,1)}_{\ti{j}_2,m_2+1,\bar{m}_ 2}
\ti{V}^{(-1,-1)}_{\ti{j}_3,m_3,\bar{m}_3} \lb \cdot \la
V^{(0,1)}_{j_1,m_1+1,\bar{m}_1}V^{(1,0)}_{j_2,m_2,\bar{m}_2+1}
V^{(-1,-1)}_{j_3,m_3,\bar{m}_3} \lb, \cr &c=
\bar{k}^{(2)}_{L}\bar{E}^{(1)}_{L}
\bar{k}^{(1)}_{R}\bar{E}^{(2)}_{R} \cdot \la
\ti{V}^{(0,1)}_{\ti{j}_1,m_1+1,\bar{m}_1}
\ti{V}^{(1,0)}_{\ti{j}_2,m_2,\bar{m}_2+ 1}
\ti{V}^{(-1,-1)}_{\ti{j}_3,m_3,\bar{m}_3} \lb \cdot \la
V^{(1,0)}_{j_1,m_1,\bar{m}_1+1}V^{(0,1)}_{j_2,m_2+1,\bar{m}_2}
V^{(-1,-1)}_{j_3,m_3,\bar{m}_3}, \cr & d=
\bar{k}^{(2)}_{L}\bar{E}^{(1)}_{L}
\bar{k}^{(2)}_{R}\bar{E}^{(1)}_{R} \cdot \la
\ti{V}^{(0,0)}_{\ti{j}_1,m_1+1,\bar{m}_1+1}
\ti{V}^{(1,1)}_{\ti{j}_2,m_2,\bar{m}_ 2}
\ti{V}^{(-1,-1)}_{\ti{j}_3,m_3,\bar{m}_3} \lb \cdot \la
V^{(1,1)}_{j_1,m_1,\bar{m}_1}V^{(0,0)}_{j_2,m_2+1,\bar{m}_2+1}
V^{(-1,-1)}_{j_3,m_3,\bar{m}_3} \lb ,}} and the $e$ and $f$ are
given by \eqn\uouoddfefm{\eqalign{& e=
\bar{k}^{(1)}_{L}\bar{k}^{(2)}_{L}
\bar{k}^{(1)}_{R}\bar{k}^{(2)}_{R} \cdot \la
\ti{V}^{(1,1)}_{\ti{j}_1,m_1,\bar{m}_1}
\ti{V}^{(1,1)}_{\ti{j}_2,m_2,\bar{m}_2}
\ti{V}^{(-1,-1)}_{\ti{j}_3,m_3,\bar{m}_3} \lb \cdot \la
V^{(0,0)}_{j_1,m_1+1,\bar{m}_1+1}V^{(0,0)}_{j_2,m_2+1,\bar{m}_2+1}
V^{(-1,-1)}_{j_3,m_3,\bar{m}_3} \lb , \cr & f=
\bar{E}^{(1)}_{L}\bar{E}^{(2)}_{L}
\bar{E}^{(1)}_{R}\bar{E}^{(2)}_{R} \cdot \la
\ti{V}^{(0,0)}_{\ti{j}_1,m_1+1,\bar{m}_1+1}
\ti{V}^{(0,0)}_{\ti{j}_2,m_2+1,\bar{m }_2+1}
\ti{V}^{(-1,-1)}_{\ti{j}_3,m_3,\bar{m}_3} \lb \cdot \la
V^{(1,1)}_{j_1,m_1,\bar{m}_1}V^{(1,1)}_{j_2,m_2,\bar{m}_2}
V^{(-1,-1)}_{j_3,m_3,\bar{m}_3} \lb.}} In the above equations
$k_{L,R}$ means the left and right-moving part of the momentum.
The value $\ti{j}$ is equal to $-j-1$ for $(+)$ type and to $j$
for $(-)$ type.

\vskip .2in

{\bf Computations of $e$ and $f$}

It is easy to calculate $e$ and $f$. The results in the $(---)$
case are \eqn\compuefa{e=f=Q^4\ti{D}_{\pm}(j_a)\cdot
D_{\mp}(j_a)\cdot
\prod_{a=1}^{3}\f{\Gamma(j_a+m_a+1)\Gamma(j_a-m_a+1)}
{\Gamma(-j_a+\bar{m}_a)\Gamma(-j_a-\bar{m}_a)},} where
$\ti{D}_{\pm}(j_a)$ is the Wick-rotated version (i.e. $b\to ib$)
of the $D_{\pm}$. On the other hand, in the $(+++)$ case we obtain
\eqn\compuefda{e=f=Q^4\ti{D}_{\pm}(-j_a-1)\cdot D_{\mp}(j_a).} It
is also straightforward to obtain the result for other cases. In
the end we obtain \finalsa.

\vskip .2in

{\bf Computations of $a,b,c$ and $d$}

Consider the $(---)$ amplitude again, because we can reproduce
other ones by the reflection relation \reflectwg. For generic
values of $(j_a,m_a,\bar{m}_a)$ we find \eqn\resfftdfz{\eqalign{&
a=d=Q^4(j-m)^2(j-\bar{m})^2\cdot \ti{D}(j_a)D(j_a)\cr &\  \ \
\cdot F(j_a,m_1,\bar{m}_1,m_2+1,\bar{m}_2+1,m_3,\bar{m}_3)
F(j_a,m_1+1,\bar{m}_1+1,m_2,\bar{m}_2,m_3,\bar{m}_3),\cr &
b=c=Q^4(j-m)^2(j-\bar{m})^2\cdot \ti{D}(j_a)D(j_a) \cr  &\  \ \
\cdot F(j_a,m_1+1,\bar{m}_1,m_2,\bar{m}_2+1,m_3,\bar{m}_3)
F(j_a,m_1,\bar{m}_1+1,m_2+1,\bar{m}_2,m_3,\bar{m}_3),}} where
$\ti{D}(j_a)$ is again the Wick-rotated version of $D(j_a)$. It
seems that this cannot be rewritten by simple functions in generic
cases; it is known that the function $F(j_a,m_a,\bar{m}_a)$ is
expressed in terms of the hypergeometric function shown in \HoA.

However, in the particular case where the function $F$ is
factorized as $F(j_a,m_a,\bar{m}_a)=f(j_a,m_a)g(j_a,\bar{m_a})$,
it is clear that $a=b=c=d$ and thus these contributions from
$a,b,c$ and $d$ in \uouofm\ are vanishing. Such examples can be
found e.g. when (i) $j_a=m_a=\bar{m}_a$ for one of $a=1,2,3$ and
when (ii) either of the four conditions in \moconxz\ is satisfied.

The results for other instanton sectors we can obtain the results
in the same way by just shifting $m_a$ and $\bar{m}_a$ by $\pm 1$
and flip some of the signs appropriately. So we omit writing down
the full expression.

\listrefs

\end